\documentclass[10pt,twocolumn,letterpaper]{article}

\usepackage{wacv}      

\usepackage{graphicx}
\usepackage{amsmath}
\usepackage{amssymb}
\usepackage{booktabs}
\usepackage{physics}
\usepackage{float}
\usepackage{amsfonts}
\usepackage{amsmath}
\usepackage{amssymb}
\usepackage{xcolor}
\usepackage{diagbox}
\usepackage{longtable}
\usepackage{booktabs}
\usepackage{multirow}
\usepackage{cite}
\usepackage[colorlinks=true,linkcolor=blue,citecolor=blue]{hyperref} %

\usepackage[capitalize]{cleveref}
\crefname{section}{Sec.}{Secs.}
\Crefname{section}{Section}{Sections}
\Crefname{table}{Table}{Tables}
\crefname{table}{Tab.}{Tabs.}

\def\wacvPaperID{*****} 
\def\confName{WACV}
\def\confYear{2024}

\begin{document}

\title{Drive as You Speak: Enabling Human-Like Interaction with Large Language Models in Autonomous Vehicles}


\author{%
  Can Cui$^{1,*}$, Yunsheng Ma$^1$, Xu Cao$^{2}$, Wenqian Ye$^{2, 3}$, Ziran Wang$^1$ \\
  1 Purdue University, West Lafayette, IN, USA, 47907 \\
  2 PediaMed.AI Lab, Shenzhen, China, 518048 \\
  3 University of Virginia, Charlottesville, VA, USA, 22903 \\
  \texttt{\{cancui\}@purdue.edu}
}
\maketitle

\begin{abstract}
   The future of autonomous vehicles lies in the convergence of human-centric design and advanced AI capabilities. Autonomous vehicles of the future will not only transport passengers but also interact and adapt to their desires, making the journey comfortable, efficient, and pleasant. In this paper, we present a novel framework that leverages Large Language Models (LLMs) to enhance autonomous vehicles' decision-making processes. By integrating LLMs' natural language capabilities and contextual understanding, specialized tools usage, synergizing reasoning, and acting with various modules on autonomous vehicles, this framework aims to seamlessly integrate the advanced language and reasoning capabilities of LLMs into autonomous vehicles. The proposed framework holds the potential to revolutionize the way autonomous vehicles operate, offering personalized assistance, continuous learning, and transparent decision-making, ultimately contributing to safer and more efficient autonomous driving technologies.
\end{abstract}

\section{Introduction}
\label{sec:intro}
Recently, Large Language Models (LLMs) have attracted significant attention. The key to their success lies in their remarkable ability to process a wide range of word-based inputs, including prompts, questions, dialogues, and vocabulary spanning diverse domains, resulting in significant and coherent textual outputs. LLMs serve as vast storehouses of abundant information and knowledge acquired from numerous texts, much like the human brain. Considering the LLM's ability to emulate the human brain functions, it prompts us to ask: could we leverage the impressive capabilities of LLMs to revolutionize the future of autonomous driving?

Imagine a situation where you're sitting in an autonomous vehicle and you desire to safely overtake another vehicle. All you have to do is utter the command: ``Overtake the vehicle in front of me." At that point, the LLMs would swiftly assess the existing conditions and safety listen, and ask questions before reasoning, providing you with informed guidance on the feasibility and recommended actions for executing the maneuver. Furthermore, in the context of fully autonomous vehicles, the LLMs' capabilities could even extend to taking charge of the vehicle and executing the instructed commands.

While LLMs have the potential to greatly enhance convenience and improve the driving experience for drivers, a significant challenge arises:  LLMs lack understanding of information about the driving environment. Unlike humans, LLMs lack the inherent ability to perceive the physical environment. In other words, these models do not possess the capability to visually perceive and interact with the world around them \cite{bender_climbing_2020}. It renders LLMs challenged in making sound decisions for the current situation, potentially leading to suboptimal outcomes or even hazardous consequences.

To address the challenge above, we present a perspective where LLMs can serve as the decision-making ``brain" within autonomous vehicles. Complementing this, various tools within the autonomous vehicle ecosystem, including the perception module, localization module, and in-cabin monitor, function as the vehicle's sensory ``eyes." This configuration enables LLMs to overcome the inherent limitation of not directly accessing real-time environmental information. By receiving processed data from the perception module, LLMs can facilitate informed decision-making, resulting in significant enhancements to the performance of the autonomous vehicle. Additionally, the vehicle's actions and controller function as its ``hands," executing instructions derived from the LLM's decision-making process. 

When comparing autonomous vehicles with and without integrated LLMs, it becomes evident that the latter offers a diverse array of compelling advantages. These advantages extend across various aspects of functionality and performance:
\begin{itemize}
\item \textbf{Language Interaction:}  LLMs enable intuitive communication between drivers and vehicles, transforming interactions from rigid commands to natural conversations.
\item \textbf{Contextual Understanding and Reasoning:} LLMs in vehicles offer enhanced contextual understanding from diverse sources like traffic laws and accident reports, ensuring decisions prioritize safety and regulation adherence.
\item \textbf{Zero-Shot Planning:} LLMs in vehicles can understand and reason about unfamiliar situations without prior exposure, allowing vehicles to navigate uncharted scenarios confidently.
\item \textbf{Continuous Learning and Personalization:} LLMs learn and adapt continuously, tailoring their assistance to individual driver preferences and improving the driving experience over time.
\item \textbf{Transparency and Trust:} LLMs can articulate their decisions in simple language, fostering a crucial bond of trust and understanding between the technology and its users.

\end{itemize}

\begin{figure*}[t]
    \centering
    \includegraphics[width=0.8\linewidth]{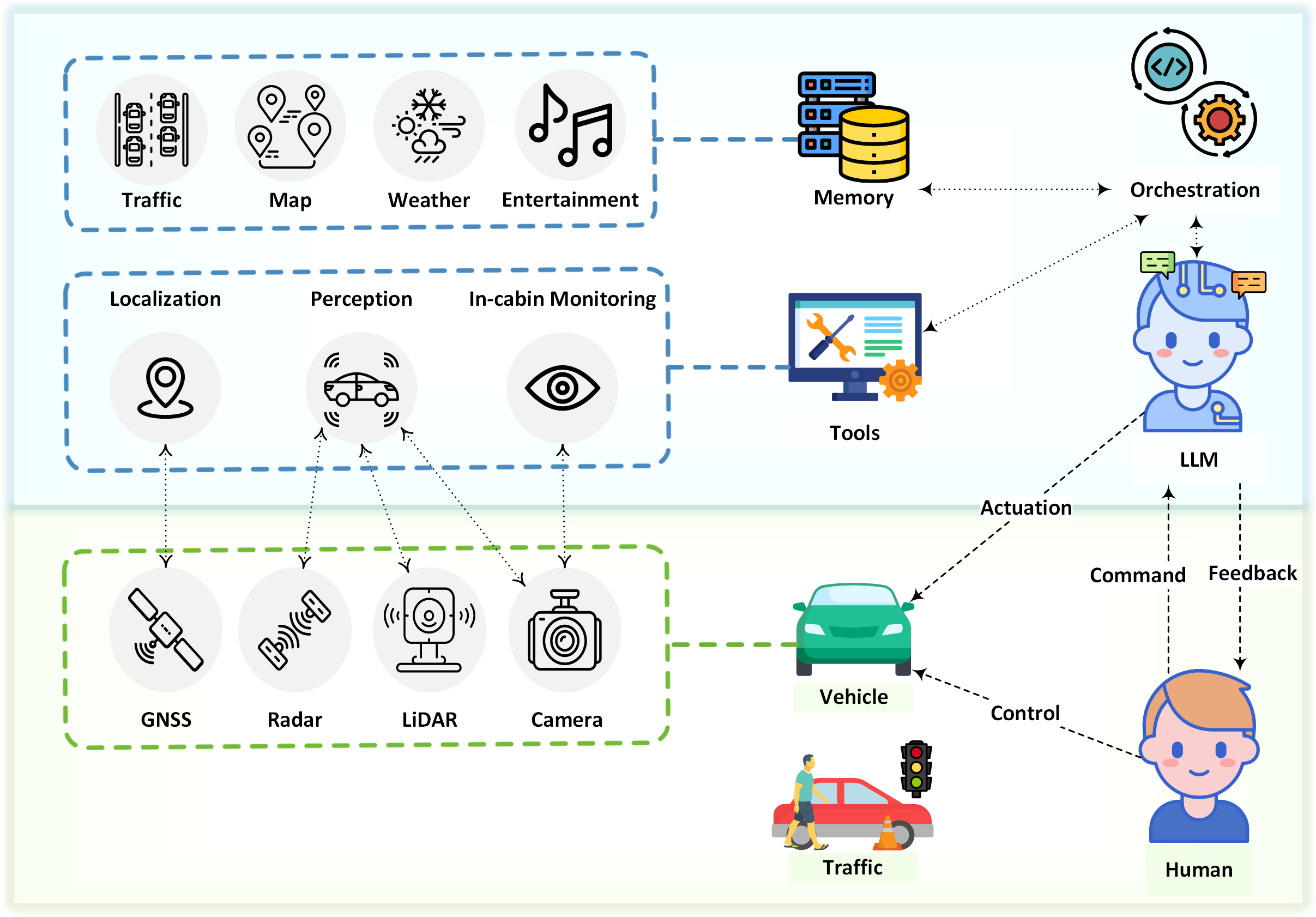}
    \caption{The human-centric LLM-integrated framework for autonomous vehicles.}
    \label{fig:framework}
\end{figure*}

\section{Perspective: the Role of LLMs in Advancing Autonomous Vehicles}
\label{sec:perspective}

As previously established in the earlier section, we've established that the LLMs serve as the ``brain" in the autonomous driving system, facilitating driver interaction and decision-making, while the useful sensory tools and actuation function as the vehicle's ``eyes" and ``hands" respectively. To be more specific, When a driver requests a particular operation, the LLM prompts the related modules to provide data that has been processed to extract relevant information from the environment. By integrating the linguistic analysis of LLMs with the processed sensory inputs from the selected modules, the LLM can then make well-informed decisions. If the command is deemed both feasible and safe based on the prior analysis, the LLMs will transmit the corresponding instructions to the vehicle's controller. This includes components such as the steering wheel, throttle pedal, braking, and other control elements, enabling them to execute the necessary operations. Alternatively, if the operation is deemed inappropriate, the LLMs will provide drivers with a detailed explanation as to why the requested action is not suitable for execution.

Revisiting the example at the beginning of this paper, when drivers issue the command to overtake the vehicle ahead, the LLMs come into play by querying the perception module for pertinent processed information. This includes details such as the distance and speed of the target vehicle, the velocity of the ego vehicle, road conditions of potential lanes, the presence of other vehicles and their distances on those lanes, and other useful navigation information from the map system. Through an analysis of the provided data and the given command, the LLMs make a decision regarding whether to execute the driver's request. If the decision is affirmative, the LLMs subsequently communicate instructions to the controller, guiding the next course of action.

Having explored this intricate interaction between LLMs and the autonomous vehicle's decision-making process, we shift our focus to a broader context and propose the concept of a human-centric LLMs integrated framework for autonomous vehicles based on our prior work of the mobility digital twin \cite{wang_mobility_2022}. As shown in \cref{fig:framework}, the physical world comprises human drivers, vehicles, and traffic objects. In the physical world, human drivers are the central agents in the physical world, sending commands and instructions to LLMs as they navigate roadways. The traffic environment contains various elements including vehicles, pedestrians, traffic lights, road conditions, and traffic cones, all of which contribute to the complexity of movement and interactions on the road. The vehicle, directed by the LLMs, operates within this ecosystem, executing the commands it receives from either drivers or LLMs through controllers and actuators.

The virtual world includes LLMs, memory, and essential tools which include the perception module, localization module, and in-cabin monitor. The perception module acquires raw input from sensors, including external cameras, LIDARs, and radars, and processes this data into a format suitable for the LLMs. The localization module employs GNSS data to determine the vehicle's precise location. Within the vehicle, the in-cabin monitor employs internal cameras, thermometers, and other sensors to vigilantly observe the in-cabin environment, preempting distractions, extreme temperatures, or uncomfortable conditions. At the core of the entire framework lies the LLMs, serving as its central intelligence. They receive commands from drivers, subsequently initiating queries to pertinent modules for related information. Furthermore, the memory section acts as a repository, storing historical operations and drivers' preferences, enabling continuous learning and enhancement for the LLMs. This repository of experiences equips the LLMs to make analogous decisions when confronted with similar situations, bolstering the system's adaptability and performance over time. the memory also houses maps and local law information, empowering the LLMs to make even wiser decisions adaptable to a variety of scenarios.

\section{Review: Can LLMs Really Do This?}
\label{sec:review}
Through a comprehensive review of both theoretical underpinnings and real-world implementations, we seek to address the fundamental question: Can LLMs really contribute to the improvement of autonomous driving by actively participating in the decision-making framework? By examining the current state of research and analyzing use cases, this section aims to provide a thorough assessment of the extent to which LLMs can bring to the landscape of human-centric autonomous driving.

\subsection{Adaptive Techniques and Human-Centric Refinements for LLMs}
Parameter-efficient fine tuning (PEFT) is a crucial technique used to adapt pre-trained language models (LLMs) to specialized downstream applications~\cite{chung_scaling_2022,fu_effectiveness_2023,hu_lora_2022,dettmers_qlora_2023,lester_power_2021}. Hu et al.~\cite{hu_lora_2022} proposed utilizing low-rank decomposition matrices to reduce the number of trainable parameters needed for fine-tuning language models. Lester et al.~\cite{lester_power_2021} explore prompt tuning, a method for conditioning language models with learned soft prompts, which achieves competitive performance compared to full fine-tuning and enables model reuse for many tasks. These PEFT techniques offer valuable tools for adapting LLMs to autonomous driving tasks.

Reinforcement Learning from Human Feedback (RLHF) \cite{stiennon_learning_2020,ouyang_training_2022,schulman_proximal_2017,bai_constitutional_2022,rafailov_direct_2023} has emerged as a key strategy for fine-tuning LLM systems to align more closely with human preferences. Ouyang et al.~\cite{ouyang_training_2022} introduce a human-in-the-loop process to create a model that better follows instructions. Bai et al.~\cite{bai_constitutional_2022} propose a method for training a harmless AI assistant without human labels, providing better control over AI behavior with minimal human input. These approaches hold significant promise for developing LLMs for autonomous driving applications, as they can contribute in two dimensions. Firstly, they can ensure that LLMs avoid making decisions that may be illegal or unethical. Secondly, these methodologies enable LLMs to continually adapt and align their decision-making processes with user preferences, enhancing personalization and trust in autonomous vehicles.

LLM-based autonomous driving applications can also benefit from advanced prompting techniques~\cite{wei_chain--thought_2022,kojima_large_2022,yao_react_2023,wang_self-consistency_2023,besta_graph_2023}. Chain-of-thought prompting~\cite{wei_chain--thought_2022} improves LLMs' ability to perform complex reasoning. Gao et al. \cite{gao_pal_2023} propose an approach that uses LLMs to read natural language problems and generate programs as intermediate reasoning steps. Yao et al. \cite{yao_react_2023} present a new prompting technique that allows LLMs to make decisions about how to interact with external APIs. These methods provide a solid foundation for the development of LLMs for autonomous driving applications with two significant benefits. Firstly, they greatly enhance LLMs' reasoning capabilities, particularly in complex, multi-step scenarios. Secondly, these techniques improve the adaptability and versatility of LLMs, key attributes for autonomous driving systems interfacing with various tools and data sources.

\subsection{Advancements in LLMs: Implications for Autonomous Driving Decision-Making}
Recent Research has shown that LLMs can perform well in most commonsense tasks \cite{bian_chatgpt_2023-1}, which means it has the potential to make wise and feasible decisions in autonomous driving scenarios. The utilization of LLMs in the context of autonomous driving presents a captivating and potentially transformative direction for research. Recent investigations have brought light on the diverse ways in which LLMs can profoundly impact the landscape of autonomous vehicles. For instance, the study conducted by \cite{nay_law_2022} highlights the promise of AI-infused with legal knowledge, offering the potential to avert legal transgressions in autonomous driving scenarios, thereby contributing to the establishment of a safer AI-driven environment. Additionally, \cite{zheng_trafficsafetygpt_2023} demonstrates that LLMs possess the capability to learn from local laws and accident reports, and effectively contribute to reducing accident rates, thus enhancing the safety of autonomous driving.

The application of LLMs to decision-making in autonomous driving is notably explored by\cite{chowdhery_palm_2022}. Their research introduces the PaLM model, demonstrating that LLMs exhibit a capacity to effectively tackle intricate reasoning tasks and, intriguingly, surpass the performance of an average human. Such a finding carries significant implications, hinting at LLMs' remarkable ability to navigate complex scenarios, make astute judgments, and potentially lay the groundwork for optimal decision-making in autonomous vehicles.

The work highlighted in \cite{park_generative_2023} demonstrates the utilization of large language models to effectively store experiences in natural language, forming a foundational approach for integrating historical data into our architecture. 

The adaptive capabilities of LLMs are showcased in various ways. \cite{kojima_large_2023} underscores LLMs' proficiency in zero-shot reasoning, enabling them to deal with novel and unfamiliar situations, a vital feature for autonomous vehicles operating in dynamic environments. The study by \cite{chung_scaling_2022} exemplifies that LLMs can be fine-tuned to exhibit enhanced performance, particularly in tasks with limited training data.

Additionally, LLMs have shown great potential in both transportation and robotics areas, as highlighted by \cite{zheng_chatgpt_2023}, and \cite{vemprala_chatgpt_2023} respectively. They reveal LLMs' prowess in tasks such as zero-shot planning and interactive conversations, even facilitating interaction with perception-action-based API libraries, an attribute that aligns with the demands of autonomous vehicles.

Furthermore, the work \cite{wang_voyager_2023} demonstrates LLMs' potential for continuous learning, which is of paramount importance for adapting to evolving road conditions and enhancing performance over time.

The investigation from \cite{driess_palm-e_2023} introduces embodied language models capable of assimilating real-world sensor data, thus bridging the gap between perception and language. This development lays the foundation for potential advancements in autonomous vehicles, where LLMs could process sensory inputs, comprehend their surroundings, and consequently make more informed decisions. Building on these insights, additional studies \cite{bian_chatgpt_2023-1}, \cite{zhu_ghost_2023}, \cite{yao_react_2023}, \cite{shinn_reflexion_2023},  and \cite{huang_instruct2act_2023} have further enriched our understanding of LLMs' capabilities, underscoring their potential in decision-making, reasoning, and synergizing reasoning and acting.


\begin{figure}[!t]
  \centering
  \includegraphics[width=\linewidth]{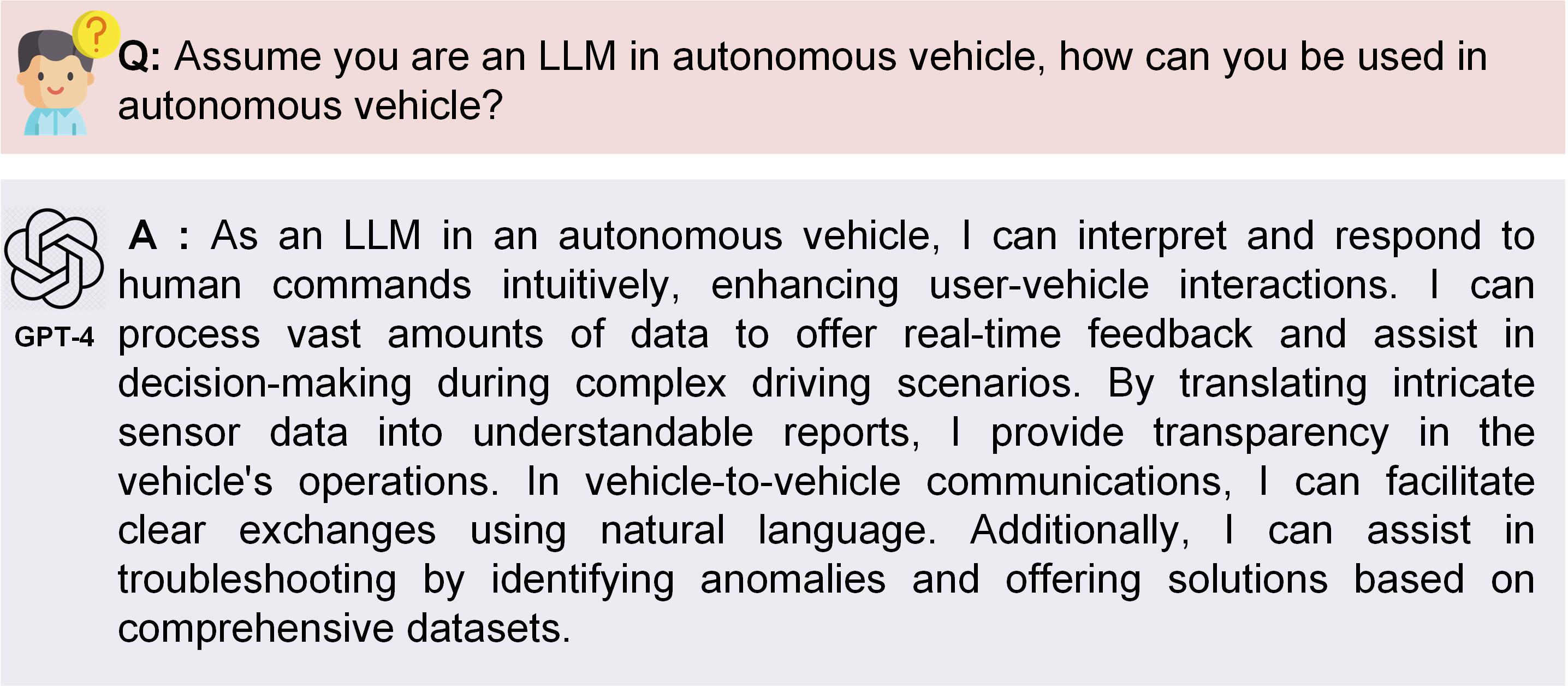}
  \vspace{0.05cm}\\
  \includegraphics[width=\linewidth]{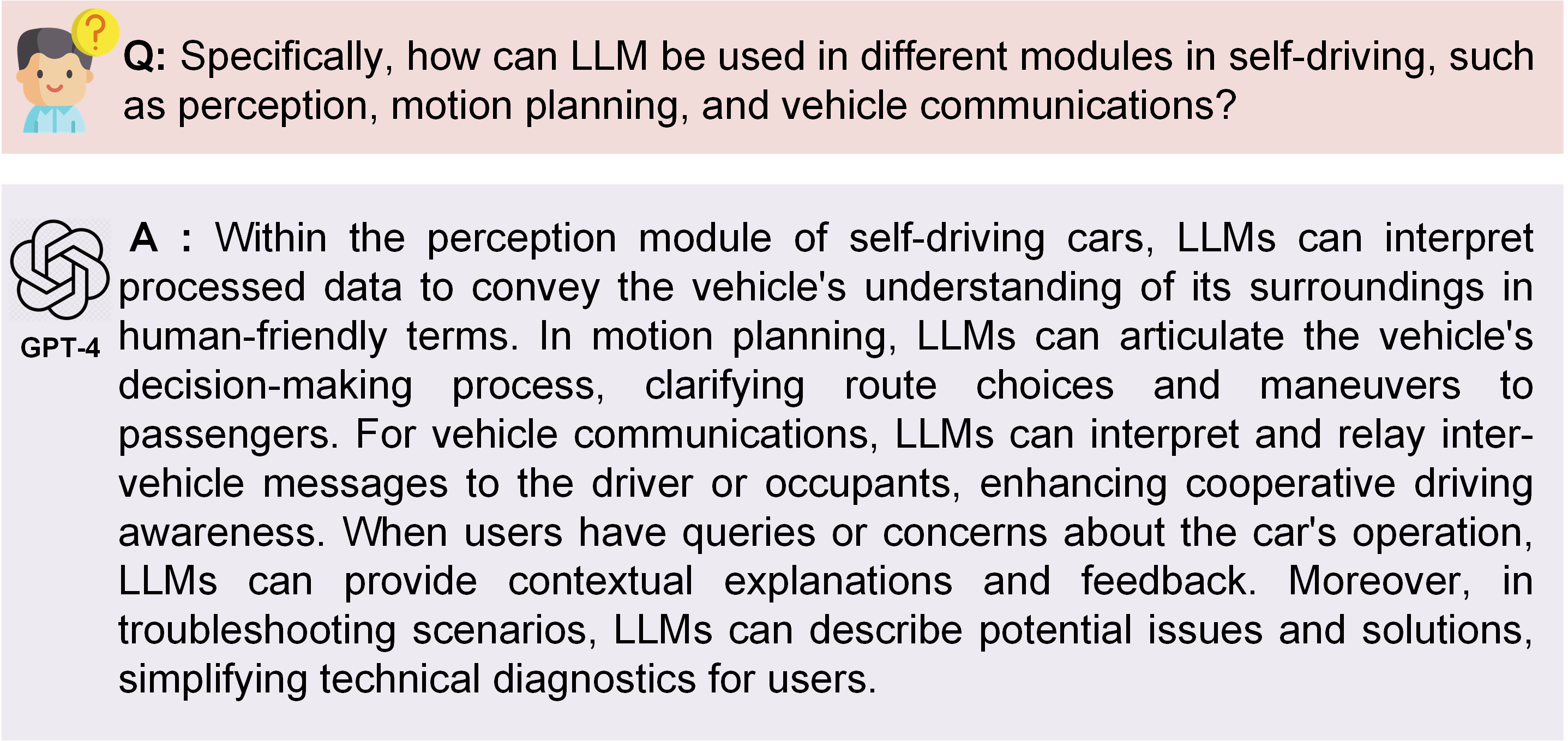}
  \caption{General Q\&A with ChatGPT-4 regarding autonomous vehicles.}
  \label{fig:general_QA}
\end{figure}

\begin{figure*}[!t]
  \centering
  \includegraphics[width=0.9\textwidth,height=19.5cm]{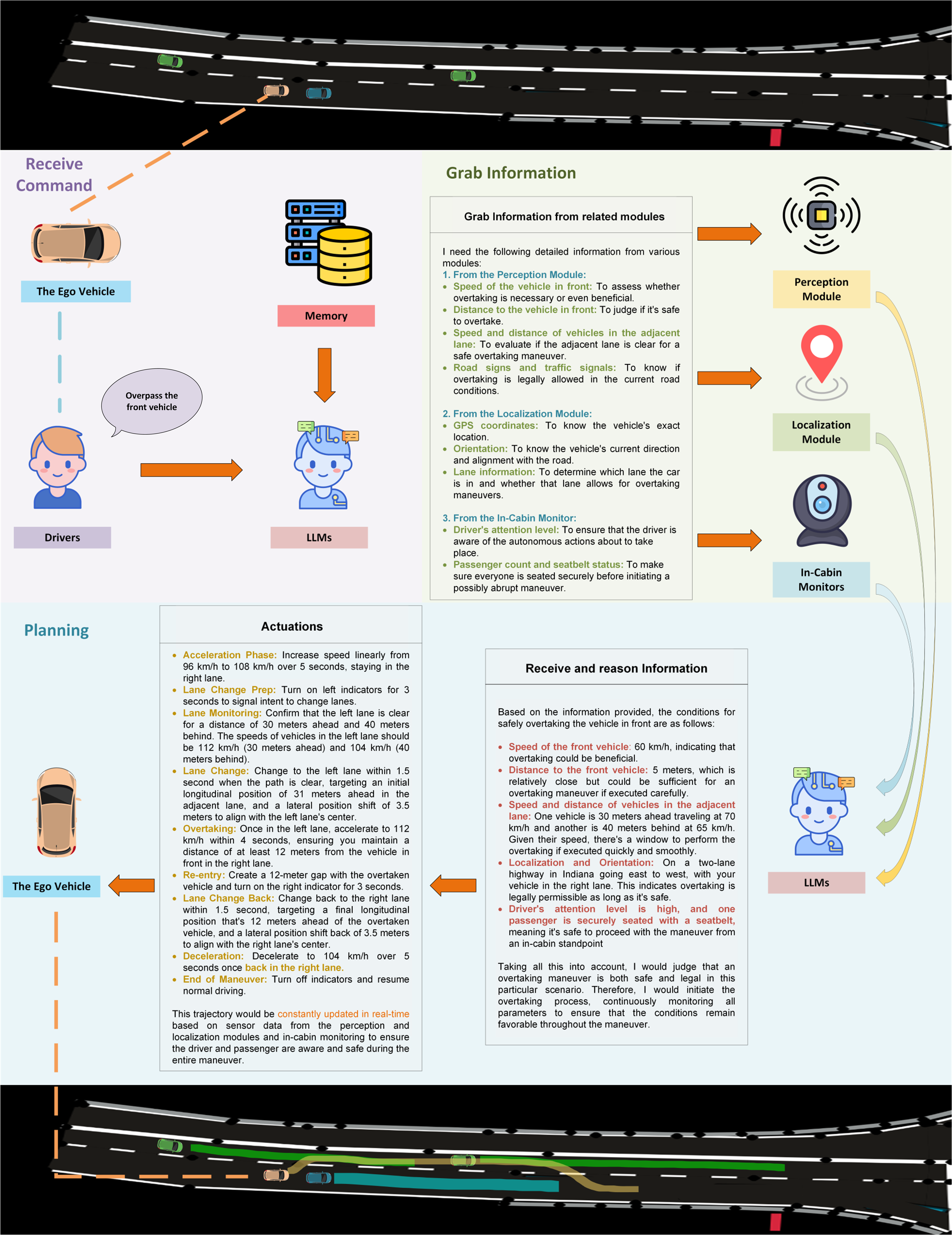}
  \caption{Experiment illustrating LLM-assisted decision-making and motion planning in a complex driving scenario. \textcolor{orange}{The Ego vehicle and its trajectory} are marked orange; \textcolor{blue}{The vehicle ahead in the current lane and its trajectory} are blue; \textcolor{green}{The vehicles in adjacent lanes and their trajectories} are green.}
  \label{fig:ex}
\end{figure*}



\section{Experiment: Decision-Making and Motion Planning with ChatGPT-4}
\label{sec:experiments}
To gain a deeper understanding of the practical capabilities of LLMs in the context of autonomous driving tasks, we embark on an insightful exploration involving real-world decision-making scenarios. This comprehensive case study serves as a compelling demonstration of how LLMs can effectively enhance autonomous vehicles by harnessing the potential of ChatGPT-4~\cite{openai_gpt-4_2023} to replicate decision-making processes. Our investigation is structured in two distinct phases. Initially, we pose autonomous dirving concept-related queries to GPT-4, which unveils its grasp of how language models can be seamlessly integrated into autonomous driving. Subsequently, we design and present genuine real-world situations to assess the decision-making proficiency of LLMs. This section covers an in-depth understanding of this case study, including a detailed conversation with GPT-4 that highlights our findings. This analysis serves to underscore the practical implications of leveraging LLMs for enhanced autonomous driving.

In our exploration with ChatGPT, we first asked some general conceptional questions regarding the LLMs in autonomous vehicles and aimed to find the true potential of LLMs in advancing the future of autonomous driving in Figure \ref{fig:general_QA}. The responses indicated a profound ability of the LLMs to bridge the interaction between the vehicle and its passengers. From the responses, it's evident that LLMs can explain complex driving scenarios, decisions made by the vehicle, and even the technical of various autonomous modules. An especially significant observation was the LLMs' strength in processing vast volumes of data, and then converting these into real-time, understandable feedback. Such feedback isn't just about driving status but relates to the core autonomous functionalities, including the perception module's utilization and the motion planning's choices. Furthermore, the model demonstrated an enhanced capacity for vehicle-to-vehicle communications and, critically, troubleshooting. This capability not only fosters trust but can also develop the user experience by explaining the complex decisions of autonomous operations.

As we can see in Figure \ref{fig:ex}, we simulated a real-world driving scenario where the autonomous vehicle is equipped with Large Language Models (LLMs) to assist in decision-making and motion planning. The vehicle was on a two-lane Indiana highway, traveling east to west at 96 km/h. It was behind another vehicle moving at the same speed but only 8 meters away, a distance less than optimal for safety. On the adjacent left lane, two other vehicles were noted: one 30 meters ahead moving at 112 km/h, and another 40 meters behind at 104 km/h. The driver was highly attentive, and one passenger was wearing a seatbelt.

The LLMs were tasked with processing this multilayered data sourced from the perception module (vehicle speeds and distances), the localization module (road and environmental conditions), and the in-cabin monitoring system (driver's attention level and safety measures like seatbelts). The LLMs formulated a comprehensive 9-step motion plan that prioritized safety while efficiently executing the driver's command to overtake the front vehicle.

In the experimental scenario, the Large Language Models (LLMs) showcased their advanced reasoning ability by not just collecting and analyzing data but also applying layers of context-sensitive reasoning. The LLMs evaluated the speeds and distances of surrounding vehicles, the driver's state of attention, and even the traffic conditions to determine the safest and most efficient trajectory for overtaking. This capability to reason in real-time, considering multiple factors dynamically, significantly contributes to road safety and operational efficacy. The LLMs didn't merely follow pre-defined rules but adapted their decision-making to the unique circumstances, highlighting their potential for enhancing the future of autonomous driving.

Additionally, the language interaction capabilities of the LLMs proved crucial for trust-building. When the driver commanded to ``overtake the vehicle in front," the LLMs assessed various factors and communicated their reasoning to the driver. This transparent interaction not only enhanced safety but also instilled greater confidence in the vehicle's autonomous capabilities.

LLMs also can access previous data and user preferences from the memory module, which allows for a more personalized driving experience. In the context of the experiment, for instance, the system could recall the driver's typical comfort levels with overtaking speeds, following distances, and lane preferences. This information could then influence how the LLMs interpret and execute a command like ``overtake the front vehicle," ensuring that the action aligns with the driver's past behavior and comfort zones. As a result, the LLMs' capacity for memory-driven personalization not only improves user satisfaction but also can contribute to safer, more predictable autonomous driving scenarios.

Another crucial advantage is enhanced transparency and trust. When the vehicle makes a complex decision, such as overtaking another vehicle on a high-speed, two-lane highway, passengers and drivers might naturally have questions or concerns. In these instances, the LLMs don't just execute the task but also articulate the reasoning behind each step of the decision-making process. By providing real-time, detailed explanations in understandable language, the LLMs demystify the vehicle's actions and underlying logic. This not only satisfies the innate human curiosity about how autonomous systems work but also builds a higher level of trust between the vehicle and its occupants.

Moreover, the advantage of ``zero-shotting" was particularly evident during the complex overtaking maneuver on a high-speed Indiana highway. Despite the LLMs not having encountered this specific set of circumstances before—varying speeds, distances, and even driver alertness—it was able to use its generalized training to safely and efficiently generate a trajectory for the overtaking action. This ensures that even in dynamic or rare scenarios, the system can make sound judgments while keeping users informed, hence building trust in autonomous technology.




\section{Conclusion}
\label{sec:conclusion}

In conclusion, our paper has provided a comprehensive framework for integrating Large Language Models (LLMs) into the ecosystem of autonomous vehicles. We highlighted how LLMs offer advanced reasoning capabilities that can make autonomous systems more flexible and responsive to complex, real-world scenarios. Additionally, by leveraging the capabilities of LLMs, we can enrich the human-vehicle interaction, providing a more reliable, intuitive, and responsive interface. Unlike traditional autonomous systems, which lack the capacity for language understanding, LLMs can handle complex requests, offer real-time feedback and comprehensive explanations, and assist in decision-making during complex or rare driving scenarios. This suggests a future where LLMs can significantly enhance efficiency, safety, and user-centric design in autonomous vehicles.

{\small
\bibliographystyle{ieee_fullname}
\bibliography{bib/ma, bib/LLM_ref}

\begin{thebibliography}{10}\itemsep=-1pt

\bibitem{bai_constitutional_2022}
Yuntao Bai, Saurav Kadavath, Sandipan Kundu, Amanda Askell, Jackson Kernion, Andy Jones, Anna Chen, Anna Goldie, Azalia Mirhoseini, Cameron McKinnon, Carol Chen, Catherine Olsson, Christopher Olah, Danny Hernandez, Dawn Drain, Deep Ganguli, Dustin Li, Eli Tran-Johnson, Ethan Perez, Jamie Kerr, Jared Mueller, Jeffrey Ladish, Joshua Landau, Kamal Ndousse, Kamile Lukosuite, Liane Lovitt, Michael Sellitto, Nelson Elhage, Nicholas Schiefer, Noemi Mercado, Nova DasSarma, Robert Lasenby, Robin Larson, Sam Ringer, Scott Johnston, Shauna Kravec, Sheer~El Showk, Stanislav Fort, Tamera Lanham, Timothy Telleen-Lawton, Tom Conerly, Tom Henighan, Tristan Hume, Samuel~R. Bowman, Zac Hatfield-Dodds, Ben Mann, Dario Amodei, Nicholas Joseph, Sam McCandlish, Tom Brown, and Jared Kaplan.
\newblock Constitutional {AI}: {Harmlessness} from {AI} {Feedback}, Dec. 2022.
\newblock arXiv:2212.08073 [cs].

\bibitem{bender_climbing_2020}
Emily~M. Bender and Alexander Koller.
\newblock Climbing towards {NLU}: {On} {Meaning}, {Form}, and {Understanding} in the {Age} of {Data}.
\newblock In {\em Proceedings of the 58th {Annual} {Meeting} of the {Association} for {Computational} {Linguistics}}, pages 5185--5198, Online, 2020. Association for Computational Linguistics.

\bibitem{besta_graph_2023}
Maciej Besta, Nils Blach, Ales Kubicek, Robert Gerstenberger, Lukas Gianinazzi, Joanna Gajda, Tomasz Lehmann, Michal Podstawski, Hubert Niewiadomski, Piotr Nyczyk, and Torsten Hoefler.
\newblock Graph of {Thoughts}: {Solving} {Elaborate} {Problems} with {Large} {Language} {Models}, Aug. 2023.
\newblock arXiv:2308.09687 [cs].

\bibitem{bian_chatgpt_2023-1}
Ning Bian, Xianpei Han, Le Sun, Hongyu Lin, Yaojie Lu, and Ben He.
\newblock {ChatGPT} is a {Knowledgeable} but {Inexperienced} {Solver}: {An} {Investigation} of {Commonsense} {Problem} in {Large} {Language} {Models}, Mar. 2023.
\newblock arXiv:2303.16421 [cs].

\bibitem{chowdhery_palm_2022}
Aakanksha Chowdhery, Sharan Narang, Jacob Devlin, Maarten Bosma, Gaurav Mishra, Adam Roberts, Paul Barham, Hyung~Won Chung, Charles Sutton, Sebastian Gehrmann, Parker Schuh, Kensen Shi, Sasha Tsvyashchenko, Joshua Maynez, Abhishek Rao, Parker Barnes, Yi Tay, Noam Shazeer, Vinodkumar Prabhakaran, Emily Reif, Nan Du, Ben Hutchinson, Reiner Pope, James Bradbury, Jacob Austin, Michael Isard, Guy Gur-Ari, Pengcheng Yin, Toju Duke, Anselm Levskaya, Sanjay Ghemawat, Sunipa Dev, Henryk Michalewski, Xavier Garcia, Vedant Misra, Kevin Robinson, Liam Fedus, Denny Zhou, Daphne Ippolito, David Luan, Hyeontaek Lim, Barret Zoph, Alexander Spiridonov, Ryan Sepassi, David Dohan, Shivani Agrawal, Mark Omernick, Andrew~M. Dai, Thanumalayan~Sankaranarayana Pillai, Marie Pellat, Aitor Lewkowycz, Erica Moreira, Rewon Child, Oleksandr Polozov, Katherine Lee, Zongwei Zhou, Xuezhi Wang, Brennan Saeta, Mark Diaz, Orhan Firat, Michele Catasta, Jason Wei, Kathy Meier-Hellstern, Douglas Eck, Jeff Dean, Slav Petrov, and Noah Fiedel.
\newblock {PaLM}: {Scaling} {Language} {Modeling} with {Pathways}, Oct. 2022.
\newblock arXiv:2204.02311 [cs].

\bibitem{chung_scaling_2022}
Hyung~Won Chung, Le Hou, Shayne Longpre, Barret Zoph, Yi Tay, William Fedus, Yunxuan Li, Xuezhi Wang, Mostafa Dehghani, Siddhartha Brahma, Albert Webson, Shixiang~Shane Gu, Zhuyun Dai, Mirac Suzgun, Xinyun Chen, Aakanksha Chowdhery, Alex Castro-Ros, Marie Pellat, Kevin Robinson, Dasha Valter, Sharan Narang, Gaurav Mishra, Adams Yu, Vincent Zhao, Yanping Huang, Andrew Dai, Hongkun Yu, Slav Petrov, Ed~H. Chi, Jeff Dean, Jacob Devlin, Adam Roberts, Denny Zhou, Quoc~V. Le, and Jason Wei.
\newblock Scaling {Instruction}-{Finetuned} {Language} {Models}, Dec. 2022.
\newblock arXiv:2210.11416 [cs].

\bibitem{dettmers_qlora_2023}
Tim Dettmers, Artidoro Pagnoni, Ari Holtzman, and Luke Zettlemoyer.
\newblock {QLoRA}: {Efficient} {Finetuning} of {Quantized} {LLMs}, May 2023.
\newblock arXiv:2305.14314 [cs].

\bibitem{driess_palm-e_2023}
Danny Driess, Fei Xia, Mehdi S.~M. Sajjadi, Corey Lynch, Aakanksha Chowdhery, Brian Ichter, Ayzaan Wahid, Jonathan Tompson, Quan Vuong, Tianhe Yu, Wenlong Huang, Yevgen Chebotar, Pierre Sermanet, Daniel Duckworth, Sergey Levine, Vincent Vanhoucke, Karol Hausman, Marc Toussaint, Klaus Greff, Andy Zeng, Igor Mordatch, and Pete Florence.
\newblock {PaLM}-{E}: {An} {Embodied} {Multimodal} {Language} {Model}, Mar. 2023.
\newblock arXiv:2303.03378 [cs].

\bibitem{fu_effectiveness_2023}
Zihao Fu, Haoran Yang, Anthony Man-Cho So, Wai Lam, Lidong Bing, and Nigel Collier.
\newblock On the {Effectiveness} of {Parameter}-{Efficient} {Fine}-{Tuning}.
\newblock In {\em {AAAI}}, volume~37, pages 12799--12807, June 2023.
\newblock Number: 11.

\bibitem{gao_pal_2023}
Luyu Gao, Aman Madaan, Shuyan Zhou, Uri Alon, Pengfei Liu, Yiming Yang, Jamie Callan, and Graham Neubig.
\newblock {PAL}: {Program}-aided {Language} {Models}.
\newblock In {\em {ICML}}, 2023.

\bibitem{hu_lora_2022}
Edward~J. Hu, Yelong Shen, Phillip Wallis, Zeyuan Allen-Zhu, Yuanzhi Li, Shean Wang, Lu Wang, and Weizhu Chen.
\newblock {LoRA}: {Low}-{Rank} {Adaptation} of {Large} {Language} {Models}.
\newblock In {\em {ICLR}}, 2022.

\bibitem{huang_instruct2act_2023}
Siyuan Huang, Zhengkai Jiang, Hao Dong, Yu Qiao, Peng Gao, and Hongsheng Li.
\newblock {Instruct2Act}: {Mapping} {Multi}-modality {Instructions} to {Robotic} {Actions} with {Large} {Language} {Model}, May 2023.
\newblock arXiv:2305.11176 [cs].

\bibitem{kojima_large_2022}
Takeshi Kojima, Shixiang~(Shane) Gu, Machel Reid, Yutaka Matsuo, and Yusuke Iwasawa.
\newblock Large {Language} {Models} are {Zero}-{Shot} {Reasoners}.
\newblock In {\em {NeurIPS}}, volume~35, pages 22199--22213, 2022.

\bibitem{kojima_large_2023}
Takeshi Kojima, Shixiang~Shane Gu, Machel Reid, Yutaka Matsuo, and Yusuke Iwasawa.
\newblock Large {Language} {Models} are {Zero}-{Shot} {Reasoners}, Jan. 2023.
\newblock arXiv:2205.11916 [cs].

\bibitem{lester_power_2021}
Brian Lester, Rami Al-Rfou, and Noah Constant.
\newblock The {Power} of {Scale} for {Parameter}-{Efficient} {Prompt} {Tuning}.
\newblock In {\em {EMNLP}}, 2021.

\bibitem{nay_law_2022}
John~J. Nay.
\newblock Law {Informs} {Code}: {A} {Legal} {Informatics} {Approach} to {Aligning} {Artificial} {Intelligence} with {Humans}.
\newblock {\em Northwestern Journal of Technology and Intellectual Property}, 20(3):309--392, 2022.

\bibitem{openai_gpt-4_2023}
OpenAI.
\newblock {GPT}-4 {Technical} {Report}, Mar. 2023.

\bibitem{ouyang_training_2022}
Long Ouyang, Jeffrey Wu, Xu Jiang, Diogo Almeida, Carroll Wainwright, Pamela Mishkin, Chong Zhang, Sandhini Agarwal, Katarina Slama, Alex Ray, John Schulman, Jacob Hilton, Fraser Kelton, Luke Miller, Maddie Simens, Amanda Askell, Peter Welinder, Paul~F. Christiano, Jan Leike, and Ryan Lowe.
\newblock Training language models to follow instructions with human feedback.
\newblock In {\em {NeurIPS}}, volume~35, pages 27730--27744, 2022.

\bibitem{park_generative_2023}
Joon~Sung Park, Joseph~C. O'Brien, Carrie~J. Cai, Meredith~Ringel Morris, Percy Liang, and Michael~S. Bernstein.
\newblock Generative {Agents}: {Interactive} {Simulacra} of {Human} {Behavior}, Aug. 2023.
\newblock arXiv:2304.03442 [cs].

\bibitem{rafailov_direct_2023}
Rafael Rafailov, Archit Sharma, Eric Mitchell, Stefano Ermon, Christopher~D. Manning, and Chelsea Finn.
\newblock Direct {Preference} {Optimization}: {Your} {Language} {Model} is {Secretly} a {Reward} {Model}, May 2023.
\newblock arXiv:2305.18290 [cs].

\bibitem{schulman_proximal_2017}
John Schulman, Filip Wolski, Prafulla Dhariwal, Alec Radford, and Oleg Klimov.
\newblock Proximal {Policy} {Optimization} {Algorithms}, Aug. 2017.
\newblock arXiv:1707.06347 [cs].

\bibitem{shinn_reflexion_2023}
Noah Shinn, Federico Cassano, Beck Labash, Ashwin Gopinath, Karthik Narasimhan, and Shunyu Yao.
\newblock Reflexion: {Language} {Agents} with {Verbal} {Reinforcement} {Learning}, June 2023.
\newblock arXiv:2303.11366 [cs].

\bibitem{stiennon_learning_2020}
Nisan Stiennon, Long Ouyang, Jeffrey Wu, Daniel Ziegler, Ryan Lowe, Chelsea Voss, Alec Radford, Dario Amodei, and Paul~F Christiano.
\newblock Learning to summarize with human feedback.
\newblock In {\em {NeurIPS}}, volume~33, pages 3008--3021, 2020.

\bibitem{vemprala_chatgpt_2023}
Sai Vemprala, Rogerio Bonatti, Arthur Bucker, and Ashish Kapoor.
\newblock {ChatGPT} for {Robotics}: {Design} {Principles} and {Model} {Abilities}, July 2023.
\newblock arXiv:2306.17582 [cs].

\bibitem{wang_voyager_2023}
Guanzhi Wang, Yuqi Xie, Yunfan Jiang, Ajay Mandlekar, Chaowei Xiao, Yuke Zhu, Linxi Fan, and Anima Anandkumar.
\newblock Voyager: {An} {Open}-{Ended} {Embodied} {Agent} with {Large} {Language} {Models}, May 2023.
\newblock arXiv:2305.16291 [cs].

\bibitem{wang_self-consistency_2023}
Xuezhi Wang, Jason Wei, Dale Schuurmans, Quoc Le, Ed Chi, Sharan Narang, Aakanksha Chowdhery, and Denny Zhou.
\newblock Self-{Consistency} {Improves} {Chain} of {Thought} {Reasoning} in {Language} {Models}.
\newblock In {\em {ICLR}}, 2023.

\bibitem{wang_mobility_2022}
Ziran Wang, Rohit Gupta, Kyungtae Han, Haoxin Wang, Akila Ganlath, Nejib Ammar, and Prashant Tiwari.
\newblock Mobility {Digital} {Twin}: {Concept}, {Architecture}, {Case} {Study}, and {Future} {Challenges}.
\newblock {\em IEEE Internet of Things Journal}, 9(18):17452--17467, Sept. 2022.

\bibitem{wei_chain--thought_2022}
Jason Wei, Xuezhi Wang, Dale Schuurmans, Maarten Bosma, Brian Ichter, Fei Xia, Ed Chi, Quoc Le, and Denny Zhou.
\newblock Chain-of-{Thought} {Prompting} {Elicits} {Reasoning} in {Large} {Language} {Models}.
\newblock In {\em {NeurIPS}}, 2022.

\bibitem{yao_react_2023}
Shunyu Yao, Jeffrey Zhao, Dian Yu, Nan Du, Izhak Shafran, Karthik Narasimhan, and Yuan Cao.
\newblock {ReAct}: {Synergizing} {Reasoning} and {Acting} in {Language} {Models}.
\newblock In {\em {ICLR}}, 2023.

\bibitem{zheng_trafficsafetygpt_2023}
Ou Zheng, Mohamed Abdel-Aty, Dongdong Wang, Chenzhu Wang, and Shengxuan Ding.
\newblock {TrafficSafetyGPT}: {Tuning} a {Pre}-trained {Large} {Language} {Model} to a {Domain}-{Specific} {Expert} in {Transportation} {Safety}, July 2023.
\newblock arXiv:2307.15311 [cs].

\bibitem{zheng_chatgpt_2023}
Ou Zheng, Mohamed Abdel-Aty, Dongdong Wang, Zijin Wang, and Shengxuan Ding.
\newblock {ChatGPT} {Is} on the {Horizon}: {Could} a {Large} {Language} {Model} {Be} {All} {We} {Need} for {Intelligent} {Transportation}?, Mar. 2023.
\newblock arXiv:2303.05382 [cs].

\bibitem{zhu_ghost_2023}
Xizhou Zhu, Yuntao Chen, Hao Tian, Chenxin Tao, Weijie Su, Chenyu Yang, Gao Huang, Bin Li, Lewei Lu, Xiaogang Wang, Yu Qiao, Zhaoxiang Zhang, and Jifeng Dai.
\newblock Ghost in the {Minecraft}: {Generally} {Capable} {Agents} for {Open}-{World} {Environments} via {Large} {Language} {Models} with {Text}-based {Knowledge} and {Memory}, June 2023.
\newblock arXiv:2305.17144 [cs].

\end{thebibliography}
}

\end{document}